\documentclass[12pt]{article}

\usepackage{amssymb,amsmath}

\begin{document}
\begin{center}
{\Large\bf \boldmath Feynman disentangling method and group
theory} 

\vspace*{6mm}
{V.S.Popov and M.A.Trusov$^a$ }\\      
{\small \it $^a$ ITEP, Moscow, Russia}      

\end{center}

\vspace*{6mm}


\begin{abstract}
The subject of this work is to apply the modified Feynman
disentangling approach to a problem of transitions in a
non-quadratic quantum-mechanical system: a singular oscillator
with a time-dependent frequency.
\end{abstract}


\vspace*{6mm}

The method of disentangling expressions, containing
non-commuting operators (FDM), suggested by Feynman in
\cite{Feynman}, gave an elegant solution of the harmonic
oscillator excitation under the arbitrary time-dependent external
force. In further developments the FDM was applied to some other
non-stationary quantum mechanical problems, see
\cite{Popov_magnet,Popov_oscillator} and \cite{Popov_UFN} for
review. And it was shown that the transition matrix elements
calculation became much simpler if the FDM was supplied by some
considerations from the representation theory of $SU(2)$ or
$SU(1,1)$ groups.

In this paper we apply the FDM to a non-quadratic system ---
singular oscillator with a variable frequency $\omega(t)$. We
obtain the self-contained analytic expressions for the transition
amplitudes between states with definite quantum numbers (at $t\to
\pm\infty$) and calculate the generating functions for transition
probabilities. The important role of representation group theory
is discussed in this context.

The problems considered in
\cite{Feynman,Popov_oscillator} can be generalized to a model of a
singular oscillator with variable frequency:
\begin{equation}
\begin{gathered}
\Hat H=\frac{1}{2}p^2+\frac{1}{2}\omega(t)^2x^2+\frac{g}{8x^2},\\
0<x<+\infty, \quad g=\text{const},\quad g>-1,\quad \hbar=m=1.
\end{gathered}
\label{1}
\end{equation}
The frequency $\omega(t)$ is an arbitrary real time function.  As
usual, we propose the boundary conditions:
\begin{equation*}
\omega(t)\to\omega_{\pm} \quad \text{at} \quad t\to\pm\infty
\end{equation*}
which allows one to define the final and initial states of the
oscillator.

It is well known that at a fixed $t$ the instantaneous spectrum of
the Hamiltonian (\ref{1}) is equidistant (see, e.g.,
\cite{Landau}):
\begin{equation}
E_n=2\omega(n+j),\qquad j=\frac{1}{2}+\frac{1}{4}\sqrt{1+g},\qquad
n=0,1,2,\dots . \label{2}
\end{equation}

We note further that the operators
\begin{equation}
J_1=\frac{1}{4\omega_+}\left(-\omega_+^2x^2+p^2\right)+\frac{g}{16x^2},\quad
J_2=\frac{1}{4}(px+xp), \quad
J_3=\frac{1}{4\omega_+}\left(\omega_+^2x^2+p^2\right)+\frac{g}{16x^2},
\label{3}
\end{equation}
satisfy the standard commutation relations of the  $su(1,1)$
algebra:
\begin{equation}
[J_1,J_2]=-iJ_3,\qquad [J_2,J_3]=+iJ_1,\qquad [J_3,J_1]=+iJ_2 ,
\label{4}
\end{equation}
and the Hamiltonian (\ref{1}) is a linear combination of operators
$J_1$ and $J_3$:
\begin{equation}
H(t)=\left(\omega_++\frac{\omega^2(t)}{\omega_+}\right)\cdot J_3 +
\left(\omega_+-\frac{\omega^2(t)}{\omega_+}\right)\cdot J_1.
\label{1a}
\end{equation}
The instantaneous eigenfunctions of the Hamiltonian (\ref{1})
realize the irreducible unitary infinite-dimensional
representation of the non-compact $su(1,1)$ algebra. The
corresponding Casimir operator (``angular momentum'' squared)
proves to be a constant and can be calculated directly:
\begin{equation}
\mathbf{J}^2=J_3^2-J_1^2-J_2^2=\frac{g-3}{16}=j(j-1) \label{5}
\end{equation}
so the weight of this representation is $j$.

For the simplest case $\omega_+=\omega_-=1$ the initial and the
final states are the eigenfunctions of the $J_3$ operator:
\begin{equation*}
J_3\psi_n=\lambda_n\psi_n,\qquad \lambda_n=n+j.
\end{equation*}
According to Ref. \cite{Popov_oscillator}, the transition
probability between initial $|m\rangle$ and final $|n\rangle$
states can be expressed in terms of the generalized Wigner
function for the irreducible representation of the $su(1,1)$
algebra with weight $j$:
\begin{equation}
w_{mn}=\left|f^{(j)}_{n+j,m+j}\right|^2,\qquad n,m=0,1,2,\dots .
\label{6}
\end{equation}
The latter can be obtained by an analytic continuation of a
standard Wigner function for the compact $su(2)$ algebra; the
details of this technique were described in the Appendix A in
\cite{Popov_appendix}. In group theory such a method is known as
the ``Weyl unitary trick'' . The probability proves to depend on
the only real parameter $\rho$, $0\le\rho<1$, which has the same
sense as in the well-known problem of transitions in a regular
oscillator (see, e.g., \cite{Perelomov, Baz}) and can be
calculated from the classical oscillator equation of motion.

A generalization of this approach to the case of unequal initial
and final frequencies $\omega_+\ne \omega_-$ is quite obvious, as,
according to (\ref{1a}), the transformation from the initial state
basis to the final state one is simply a unitary rotation around
the axis 2, i.e., an element of the quasi-unitary group $SU(1,1)$.
Omitting intermediate calculations (all details can be found in
\cite{Trusov}), we derive the following expression for the
transition probabilities between states $|m,\omega_-\rangle$ and
$|n,\omega_+\rangle$ (see Eq. (13) from Ref.
\cite{Popov_oscillator}, for comparison):
\begin{equation}
w_{mn}=\frac{L!}{((L-S)!)^2S!}\frac{\Gamma(L+2j)}{\Gamma(S+2j)}\rho^{L-S}(1-\rho)^{2j}
\left[{}_2F_1(-S,L+2j;L-S+1;\rho) \right]^2, \label{7}
\end{equation}
where $j$ is defined in (\ref{2}) and
\begin{equation*}
L=\max(m,n),\qquad S=\min(m,n),\qquad L-S=|m-n|.
\end{equation*}
The formula (\ref{7}) furnishes the ultimate answer to the problem
given.

The Gauss hypergeometric function ${}_2F_1$ in (\ref{7}) has its
first argument being integer and negative (or zero), so it reduces
to the Jacobi polynomial. Making necessary transformations, we
obtain from (\ref{7}):
\begin{equation}
\begin{gathered}
w_{mn}=\frac{m!}{n!}\frac{\Gamma(n+2j)}{\Gamma(m+2j)}\rho^{n-m}(1-\rho)^{2j}
\left[P_m^{(n-m,2j-1)}(1-2\rho)\right]^2, \quad n\ge m,\\
w_{mn}=\frac{n!}{m!}\frac{\Gamma(m+2j)}{\Gamma(n+2j)}\rho^{m-n}(1-\rho)^{2j}
\left[P_n^{(m-n,2j-1)}(1-2\rho)\right]^2, \quad m\ge n,
\end{gathered}
\label{7a}
\end{equation}
which coincides with the standard quantum-mechanical result for
the transitions in the time-dependent singular oscillator obtained
by means of the Schr\"odinger equation solution (see \cite{Manko}
and references therein).

The last point to discuss is the generating functions for the
transition probabilities of the singular oscillator. Using the
expressions (\ref{7a}), one can derive the following relation
(see \cite{Trusov} for details):
\begin{equation}
G(u,v)=\sum_{m,n} w_{mn} u^m v^n=\frac{\nu^{2j}}{1-uv\cdot\nu^2},
\quad |u|,|v|<1 \label{9a}
\end{equation}
where
\begin{equation}
\nu=\frac{2(1-\rho)}{1-\rho(u+v)+uv+\sqrt{\left[1-\rho(u+v)+uv\right]^2-4uv(1-\rho)^2}}
\label{9b}
\end{equation}
For the limiting case of a regular oscillator ($g=0$, i.e.
$j=3/4$) the expression (\ref{9a}) corresponds to the formulas
from \cite{Husimi} for generating functions of odd transitions.

The formulas (\ref{9a}) and (\ref{9b})  are quite convenient to
compute various operator mean-values over the transition
probability distributions. In particular, for the adiabatic
invariant $I=\langle H\rangle/2\omega$ one obtains
\begin{equation}
\frac{I_+}{I_-}=\frac{\langle n+j
\rangle}{m+j}=\frac{1+\rho}{1-\rho} \label{10}
\end{equation}
for transitions from an arbitrary initial level $m$ .

To summarize, we note that in our paper the transition
probabilities of the singular oscillator have been calculated for
an arbitrary frequency $\omega(t)$. To solve the problem, the
modified FDM was applied and the representation theory for
non-compact $su(1,1)$ algebra has been used. The final result for
the transition amplitudes has been presented in a self-contained
form, which is rather convenient for further applications. The
expressions for the generating functions for transition
probabilities have been derived and the adiabatic invariant
variation at the oscillator evolution has been calculated.

\section*{Acknowledgements}

This work was partially supported by the Russian Foundation for
Basic Research (grant No. 07-02-01116) and by the Ministry of
Science and Education of the Russian Federation (grant No. RNP
2.1.1. 1972). One of the authors (M.A.T.) also thanks for partial
support the President Grant No. NSh-4961.2008.2 and the President
Grant No. MK-2130.2008.2 .


\begin{thebibliography}{99}

\bibitem{Feynman} R. P. Feynman, Phys. Rev. \textbf{84}, 108 (1951).

\bibitem{Popov_magnet} V. S. Popov, Zh. Eksp. Teor. Fiz. \textbf{35}, 985 (1958).

\bibitem{Popov_oscillator} V. S. Popov, Phys. Lett. \textbf{A342},
281 (2005).



\bibitem{Popov_UFN} V. S. Popov, Phys. Usp. \textbf{50}, 1217
(2007).

\bibitem{Landau} L. D. Landau and E. M. Lifshitz, \textit{Quantum Mechanics: Non-Relativistic
Theory} (Pergamon Press, 1977).

\bibitem{Popov_appendix} V.S. Popov, JETP \textbf{101}, 817 (2005).

\bibitem{Trusov} M. A. Trusov, in preparation.

\bibitem{Perelomov} V. S. Popov and A. M. Perelomov,
Zh. Eksp. Teor. Fiz. \textbf{57}, 1684 (1969); \textbf{56}, 1375
(1969).

\bibitem{Baz} A. I. Baz, Y. B. Zeldovich, and A. M. Perelomov,
\textit{Scattering, Reactions and Decay in Nonrelativistic Quantum
Mechanics} (Israel Program for Scientific Translations, Jerusalem,
1969).

\bibitem{Manko} I. A. Malkin and V. I. Man'ko,
\textit{Dynamical Symmetries and Coherent States of Quantum
Systems} (Nauka Publishers, Moscow, 1979) [in Russian].

\bibitem{Husimi} K. Husimi, Prog. Theor. Phys. \textbf{9}, 381
(1953).

\end{thebibliography}
\end{document}